\documentclass[a4paper,11pt]{article}
\usepackage{pos}

\usepackage{braket}
\usepackage{subfigure}

\newcommand{\ord}{\mathcal{O}}
\newcommand{\Op}{\mathcal{O}}

\newcommand{\dd}{\mathrm{d}}

\newcommand{\fig}{Figure }
\newcommand{\Fig}{Figure }
\newcommand{\tab}{table }

\newcommand{\twopt}{{2\mathrm{pt}}}

\newcommand{\fourpt}{{4\mathrm{pt}}}

\newcommand{\EE}{\mathrm{E}}
\newcommand{\MM}{\mathrm{M}}

\title{The hadronic tensor from four-point functions on the lattice}

\author*[a,b]{Christian Zimmermann}
\author[a]{Terrence Draper}
\author[c]{Jian Liang}
\author[a,b]{Keh-Fei Liu}
\author[d,e,f]{Raza Sabbir Sufian}
\author[a,b]{Bigeng Wang}

\affiliation[a]{Department of Physics and Astronomy, University of Kentucky, Lexington, KY 40506, USA.}

\affiliation[b]{Nuclear Science Division, Lawrence Berkeley National Laboratory, Berkeley, CA 94720, USA.}

\affiliation[c]{State Key Laboratory of Nuclear Physics and Technology, Institute of Quantum Matter, South China Normal University, Guangzhou 510006, China.}

\affiliation[d]{Department of Physics, New Mexico State University, Las Cruces, NM 88003, USA.}
\affiliation[e]{RIKEN-BNL Research Center, Brookhaven National Laboratory, Upton, NY 11973, USA.}
\affiliation[f]{Physics Department, Brookhaven National Laboratory, Upton, NY 11973, USA.}

\emailAdd{chrzim@lbl.gov}

\abstract{The hadronic tensor is the central non-perturbative object in the calculation of the cross section of lepton-hadron interactions like neutrino-nucleon scattering. It is usually parameterized in terms of structure functions, which encode all necessary information for all kinematic regions. Moreover, the structure functions can be factorized in terms of parton distribution functions (PDFs) and contains information on hadron resonances. On the lattice, we can calculate the corresponding matrix element of two quark-bilinear currents with a relative Euclidean time separation. The reconstruction of the hadronic tensor in Minkowski space requires appropriate dealing with the corresponding inverse problem. 
In our current work, we extend previous calculations on the nucleon by considering a much larger range of momentum transfers, which is inevitable in the context of structure functions. This can be achieved by using stochastic sources, which allows us to calculate the required four-point functions in a broad kinematic region.
We employ a clover fermion ensemble at pion mass $m_\pi = 223~\mathrm{MeV}$ and lattice spacing $a=0.085~\mathrm{fm}$. 
In these proceedings, we will give an overview of our simulation and present some first preliminary results.}

\FullConference{The 42nd International Symposium on Lattice Field Theory (LATTICE2025)\\
2-8 November 2025\\
Tata Institute of Fundamental Research, Mumbai, India\\}

\begin{document}
\maketitle

\section{Introduction}

The hadronic tensor, defined as the hadronic matrix elements of the commutator of two currents, represents a key ingredient in the description of hadron-lepton interactions. In the case of two electromagnetic currents, it is directly linked to the scattering cross section of deep inelastic scattering (DIS) of electrons and muons, which has been well known for a long time. Moreover, for two axial-vector currents, it is relevant for neutrino-nucleon scattering. This process has recently gained a lot of interest, since it is essential in the context of neutrino detecting experiments \cite{Kronfeld:2019nfb,Ruso:2022qes}.

As are many other hadronic quantities, the hadronic tensor is a non-perturbative object. Hence, the only way for \textit{ab-initio} determinations is given in the framework of lattice QCD \cite{Liu:1993cv,Liu:1999ak}. Existing lattice studies on the hadronic tensor mainly targeted the resonance region considering explicitly contributions from intermediate resonance states \cite{Liang:2019frk,Liang:2023uai}. This is different in the current study where we aim to address the deep inelastic region in order to extract the structure functions. These encode all information to describe scattering processes like deep inelastic scattering (DIS) and can be factorized in terms of parton distribution functions (PDFs). In contrast to earlier studies, this study requires access to a much broader kinematic range. Structure functions have been studied on the lattice in the past using the Feynman-Hellman method \cite{Chambers:2017dov}. In the current project, we attempt to do a more direct calculation by evaluating four-point functions.

In these proceedings we lay out our strategy how to deal with the hadronic tensor in Euclidean spacetime and describe our techniques we use to calculate the required four-point functions for the case of an unpolarized nucleon. Afterwards, we show some very first results.

\section{The hadronic tensor in Euclidean space time}

The hadronic tensor $W^{\MM}_{\mu\nu}(p,q)$ can be defined by the hadronic matrix element of the commutator of two currents for an external hadron momentum $p$ \cite{Workman:2022ynf}:

\begin{align}
W^{\MM}_{\mu\nu}(p,q,s) 
:= 
\frac{1}{4\pi} \int \dd^4 z\ e^{iqz} \bra{p,s} 
\left[ J_{\mu}(z), J_{\nu}(0) \right] \ket{p,s} \,,
\label{eq:Wdef}
\end{align}
where $J_{\mu}(x)$ can be an electromagnetic current or an axial vector current. The latter is, in particular, relevant in the context of neutrino-nucleon scattering. In our notation, $s$ denotes the hadron's spin. For the remainder of this work, we consider unpolarized matrix elements, i.e.\ average over helicity states. Moreover, we restrict ourselves to vector currents. In this case, the hadronic tensor can be decomposed in terms of two structure functions $F_1$ and $F_2$, which depend on $Q^2 = -q^2$ and $x = Q^2/(2p\cdot q)$ (Bjorken-$x$):

\begin{align}
W^{\MM}_{\mu\nu}(p,q) 
= 
\left( \frac{q_{\mu} q_{\nu}}{q^2} - g_{\mu\nu} \right) 
F_1(x,Q^2) 
+ 
\left( p_{\mu} - \frac{p\cdot q}{q^2} q_{\mu} \right) 
\left( p_{\nu} - \frac{p\cdot q}{q^2} q_{\nu} \right)
F_2(x,Q^2) \,.
\label{eq:SF}
\end{align} 
Since we work in Euclidean spacetime on the lattice, a direct calculation of \eqref{eq:Wdef} on the lattice is not possible. Instead we are forced to skip the corresponding integration in time direction and evaluate the following quantity \cite{Liu:1993cv,Liu:1999ak}:

\begin{align}
W^{\EE}_{\mu\nu}(\vec{p},\vec{q},\tau) 
:= 
\frac{1}{4E_{\vec{p}}} \int \dd^3 z\ e^{-i\vec{q}\cdot\vec{z}} 
\left.\bra{p} J_{\mu}(z) J_{\nu}(0) \ket{p}\right|_{\tau=z^4}
=
\int_0^\infty \frac{\dd \nu}{2E_{\vec{p}}} e^{-\nu\tau} W^{\MM}_{\mu\nu}(p,q) \,,
\label{eq:WEdef}
\end{align}
which is connected to the original hadronic tensor by a Laplace transform. A Euclidean version of the decomposition in terms of structure functions \eqref{eq:SF} is given by (we skip the arguments for brevity):

\begin{align}
W^{\EE}_{00}
&=
m^2 A
+ E^2 B
+ \frac{\partial^2}{\partial \tau^2} C
- 2E\frac{\partial}{\partial \tau} D
\,,\nonumber\\
W^{\EE}_{0j} = W^{\EE}_{j0}
&=
p_j E B
- q_j \frac{\partial}{\partial\tau} C
+ \left( q_j E - p_j \frac{\partial}{\partial \tau} \right) D
\,,\nonumber\\
W^{\EE}_{jk}
&=
- m^2 \delta_{jk} A
+ p_j p_k B
+ q_j q_k C
+ \left( p_j q_k + q_j p_k \right) D \,,
\label{eq:WEesys}
\end{align}
where the Euclidean structure functions $A$, $B$, $C$, and $D$ depend on $\vec{q}^2$, $\vec{p}\cdot\vec{q}$, the hadron energy $E_{\vec{p}}$ and $\tau$. Notice that, in contrast to \eqref{eq:SF}, we have four structure functions, because we did not take into account constraints by the Ward identity $q^\mu W^{\MM}_{\mu\nu} = 0$. The reason is that this would involve more factors of $\nu$, which would turn into derivatives $\partial_\tau$ after the Laplace transform so that we would end up with terms involving the fourth derivative w.r.t.\ $\tau$, which is computationally inconvenient. Instead we add these constraints by the Ward identity by adding the following equations to the system of equations \eqref{eq:WEesys}:

\begin{align}
0
&=
\left( \vec{p}\cdot\vec{q} + E\frac{\partial}{\partial \tau} \right) B + 
\left( \vec{q}^2 - \frac{\partial^2}{\partial \tau^2} \right) D
\,,\nonumber\\
0
&=
\left( \vec{p}\cdot\vec{q} + E\frac{\partial}{\partial \tau} \right) D + 
\left( \vec{q}^2 - \frac{\partial^2}{\partial \tau^2}\right) C - m^2 A \,.
\label{eq:Ward}
\end{align}
The relation to the original structure functions $F_{1,2}$ reads:

\begin{align}
\int_0^\infty \frac{\dd \nu}{2E_{\vec{p}}} e^{-\nu\tau} F_{1}(x,Q^2) 
&= 
- m^2 A(\vec{q}^2, \vec{p}\cdot\vec{q}, E_{\vec{p}}, \tau) 
\,,\nonumber\\
\int_0^\infty \frac{\dd \nu}{2E_{\vec{p}}} e^{-\nu\tau} \frac{2x}{Q^2} F_{2}(x,Q^2) 
&= 
B(\vec{q}^2, \vec{p}\cdot\vec{q}, E_{\vec{p}}, \tau) \,,
\label{eq:F12inv}
\end{align}
where $x$ and $Q^2$ are functions:

\begin{align}
Q^2 = \vec{q}^2 - \nu ^2 \,,
\qquad
x = \frac{\vec{q}^2 - \nu ^2}{2E_{\vec{p}}\nu - 2\vec{p}\cdot\vec{q}} \,.
\label{eq:Qxdef}
\end{align}
The task of determining the structure functions $F_1$ and $F_2$ from equation \eqref{eq:F12inv} represents an inverse problem. Notice that the integral \eqref{eq:F12inv} involves several different physical regions: First of all, for the hadronic tensor being non-zero, it is required to have $W^2 = (p+q)^2 \ge m^2$ in order to have on-shell final states. 
This leads to an implicit lower integration limit of $\nu_{\mathrm{min}} = E_{\vec{p}+\vec{q}}-E_{\vec{p}}$, with $E_{\vec{q}}:=\sqrt{\vec{q}^2+m^2}$. 
Moreover, once $\nu > |\vec{q}|$, we leave the scope of deep inelastic scattering (this region might be suppressed by the exponential though). Therefore, it might also be difficult to reconstruct the functions $A$ and $B$ from experimental data in order to compare the lattice data and experimental results at the Euclidean level. The situation is even more involved for $\vec{p}=\vec{0}$, since the integration range in $\nu$ barely covers the deep inelastic region which can be defined as $\sqrt{W^2_\mathrm{min}+(\vec{p}+\vec{q})^2}-E_{\vec{p}} < \nu < \sqrt{\vec{q}^2-Q^2_{\mathrm{min}}}$. Taking conservative bounds $Q^2_{\mathrm{min}} = 2~\mathrm{GeV}^2$ and $W^2_{\mathrm{min}} = 6~\mathrm{GeV}^2$ and assuming a nucleon mass $m\approx 1~\mathrm{GeV}$, this condition cannot be fulfilled unless $|\vec{q}|\gtrsim 4~\mathrm{GeV}$. Taking less conservative bounds like those given for experimental results \cite{Whitlow:1991uw}, one still needs $|\vec{q}|\gtrsim 1.5~\mathrm{GeV}$. In order to have a large enough contribution from reliably deep inelastic kinematics at $\vec{p}=\vec{0}$, one would need data from very large $|\vec{q}|$ only, where we expect lattice artifacts to be large. Therefore, we postpone attempts to deal with the inverse problem and comparisons with experimental data until our simulation also includes $\vec{p}\neq\vec{0}$, where the deep inelastic window is much broader, even at small and medium $|\vec{q}|$.

\section{Lattice calculation of four-point functions}

The core of the calculation of the hadronic tensor is the evaluation of the two-current matrix element in \eqref{eq:WEdef}. This matrix element can be expressed in terms of a Euclidean four-point correlation function as follows:

\begin{align}
\left.\bra{p} J_j(z) J_{j^\prime}(0) \ket{p}\right|_{\tau=z^4}
=
2E_{\vec{p}} V e^{-E_{\vec{p}}|\tau|}
\left.
\frac
{C_{\fourpt}^{jj^\prime}(\vec{p},\vec{z};t,\tau_0,\tau_0+\tau)}
{C_{\twopt}(\vec{p},t)}
\right|_{0\ll\tau_0,\tau_0+\tau\ll t} \,,
\label{eq:ratio}
\end{align}
where

\begin{align}
C_{\fourpt}^{jj^\prime}(\vec{p},\vec{z};t,\tau_1,\tau_2) 
:=
\left\langle
\Gamma\ 
\Op(\vec{p},t_0+t)\ 
J_j(\vec{x}+\vec{z},t_0+\tau_1)\ 
J_{j^\prime}(\vec{z},t_0+\tau_2)\ 
\overline{\Op}(\vec{p},t_0)
\right\rangle
\label{eq:4ptdef}
\end{align}
is the four-current correlation function, and $C_{\twopt}$ is the usual two-point function. $E_{\vec{p}}$ denotes the energy for a given momentum $\vec{p}$ and $V$ is the spatial volume. The interpolating operators $\Op$ and $\overline{\Op}$ are considered to respectively annihilate and create the hadron of interest, in our case a nucleon. The spinor matrix $\Gamma$ projects onto positive parity and averages w.r.t.\ the nucleon's polarization (unpolarized matrix elements). The position $\vec{x}$ and the timeslice $t_0$ are arbitrary because of translational invariance\footnote{up to limitations given by open boundary conditions of the employed ensemble.}. 


Our goal is to evaluate the matrix element \eqref{eq:ratio} in a maximally broad kinematic range, so that we have enough data points available to tackle the inverse problem. To this end, we evaluate the expression in \eqref{eq:4ptdef} for all possible vectors $\vec{z}$. A procedure for this kind of calculation has been worked out in \cite{Bali:2021gel} for the case of $\tau = 0$, i.e. the two currents being located at the same timeslice. This can be readily generalized to the required setup of our present calculation, where $\tau \neq 0$.

\begin{figure}
\includegraphics[scale=0.95]{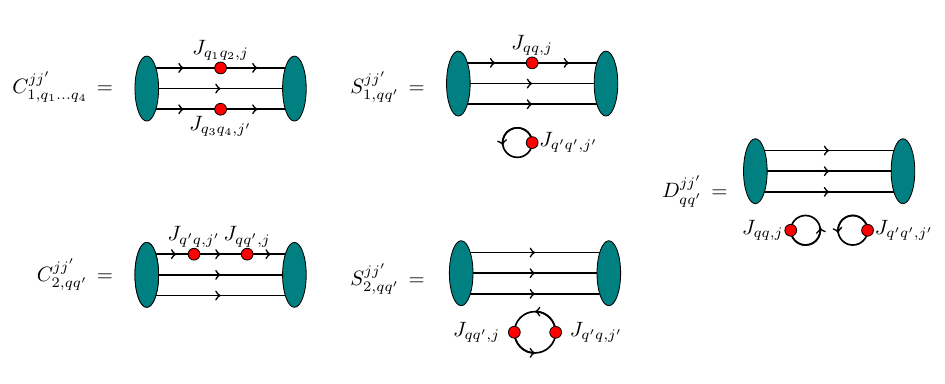}
\caption{Five types of Wick contractions contributing to the four-point function in \eqref{eq:4ptdef} for the case of baryons. The explicit expressions of the diagrams $C_1$, $C_2$, and $S_1$ depends on the quark flavor of the currents.\label{fig:wick}}
\end{figure}
The four-current correlation function decomposes in terms of Wick contractions. For the case of the nucleon, there are five kinds of contractions, which are depicted in \fig\ref{fig:wick}. The exact contribution to matrix element depends on the flavor content of the currents $J_j$ and $J_{j^\prime}$. In case of the hadronic tensor for deep inelastic scattering, we have to consider the electromagnetic current (we restrict to the case of two degenerate quark flavors only):

\begin{align}
J_{\mu}^{\mathrm{em}}(x) 
:= 
\frac{2}{3} \bar{u}(x) \gamma_\mu u(x)
-\frac{1}{3} \bar{d}(x) \gamma_\mu d(x) \,.
\label{eq:Jelm}
\end{align}
For the proton, we obtain the following decomposition:

\begin{align}
&\bra{p} J_\mu^{\mathrm{em}}(x)\ J_\nu^{\mathrm{em}}(y) \ket{p}
\propto
\frac{4}{9} C_{1,uuuu}^{jj^\prime}(x,y)
-\frac{2}{9} C_{1,uudd}^{jj^\prime}(x,y)
-\frac{2}{9} C_{1,uudd}^{j^\prime j}(y,x)
\nonumber\\
&+
\frac{4}{9} C_{2,uq}^{jj^\prime}(x,y)
+\frac{4}{9} C_{2,uq}^{j^\prime j}(y,x)
+\frac{1}{9} C_{2,dq}^{jj^\prime}(x,y)
+\frac{1}{9} C_{2,dq}^{j^\prime j}(y,x)
+\frac{2}{9} S_{1,uq}^{jj^\prime}(x,y)
\nonumber\\
&+\frac{2}{9} S_{1,uq}^{j^\prime j}(y,x)
-\frac{1}{9} S_{1,dq}^{jj^\prime}(x,y)
-\frac{1}{9} S_{1,dq}^{j^\prime j}(y,x)
+\frac{5}{9} S_{2,qq}^{jj^\prime}(x,y)
+\frac{1}{9} D_{qq}^{jj^\prime}(x,y) \,.
\label{eq:proton-elm-wick}
\end{align}
Throughout this work, we restrict ourselves to connected contributions only. It has been found in past studies that the flavor combinations $C_{1,uuuu}$ and $C_{1,uudd}$ are of similar size \cite{Bali:2021gel}. Hence, we expect a wide cancellation of the corresponding contributions in \eqref{eq:proton-elm-wick} due to the charge factors. Therefore, only $C_2$-type contractions remain as relevant connected contribution.

\begin{table}
\begin{center}
\begin{tabular}{ccccccccc}
\hline
\hline
id & $\beta$ & $a[\mathrm{fm}]$ & $L^3\times T$ & $m_\pi[\mathrm{MeV}]$ & $m_K[\mathrm{MeV}]$ & $m_\pi L$ & $L[\mathrm{fm}]$ & \# conf \\
\hline
S100 & $3.4$ & $0.085$ & $32^3\times 128$ & $223$ & $476$ & $3.1$ & $2.7$ & $984$ \\
\hline
\hline
\end{tabular}
\end{center}
\caption{Details on the CLS ensemble that has been used for this study of the hadronic tensor \cite{Bruno:2014jqa}.\label{tab:ens}}
\end{table}

Our efforts of calculating the two-current matrix elements in \eqref{eq:ratio} are part of a much more general project, since four-point functions provide access to a variety of physically relevant quantities, e.g., parton distribution functions (PDFs) \cite{Ma:2017pxb,Ji:2020ect} or double parton distributions (DPDs) \cite{Bali:2021gel}.

The calculation is carried out using gauge ensembles generated by the CLS collaboration \cite{Bruno:2014jqa}. These employ the L\"uscher-Weiss gauge action together with $\ord(a)$-improved $n_f=2+1$ Sheikholeslami-Wohlert fermions. For this first study of the direct calculation of structure functions using four-point functions, we start with the ensemble S100, which is a $32^3\times 128$-lattice, with lattice spacing $a=0.085~\mathrm{fm}$. More details are listed in \tab\ref{tab:ens}. 

As stated before, we consider only contributions of the $C_2$ diagram for now. For the corresponding evaluation, we employ point sources at the nucleon source, sequential sources at the nucleon sink, as well as time-local stochastic  sources, with $N_{\mathrm{stoch}} = 96$ sources per timeslice. The ensemble has open boundary conditions in the time direction, so we set the source time to $t_0=T/2=64a$ in order to avoid the boundary. The calculation is carried out for all possible combinations of $0\le\tau_1,\tau_2\le t$, where we use three different source-sink-separations $t\in\{8a,10a,12a\}$. Currently, our simulation includes $890$ configurations (1 nucleon point source). Moreover, we use Gaussian smearing for the nucleon source and sink. At the moment, our simulation is done only for nucleon momentum $\vec{p}=\vec{0}$.


\section{First results}

\begin{figure}
\begin{center}
\subfigure[{\parbox[t]{5cm}{$W^{\EE,u}_{44}$, $\bar{\tau}$-dependence, $\vec{q}=(2,3,4)2\pi/L$}\label{fig:W44utaubar}}]{
\includegraphics[scale=.47, clip, trim=0.4cm 0.5cm 0.4cm 1.15cm]{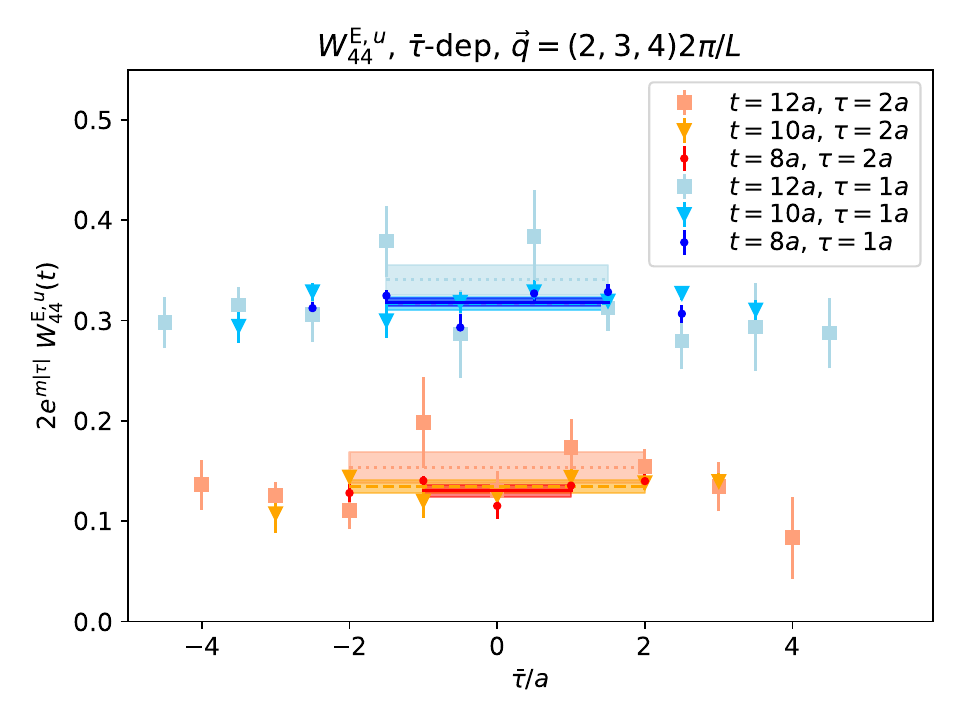}
}
\subfigure[{\parbox[t]{4cm}{$w^{\EE,d}_{44}$, $\tau$-dependence}\label{fig:W44dtau}}]{
\includegraphics[scale=.47, clip, trim=0.4cm 0.5cm 0.4cm 1.15cm]{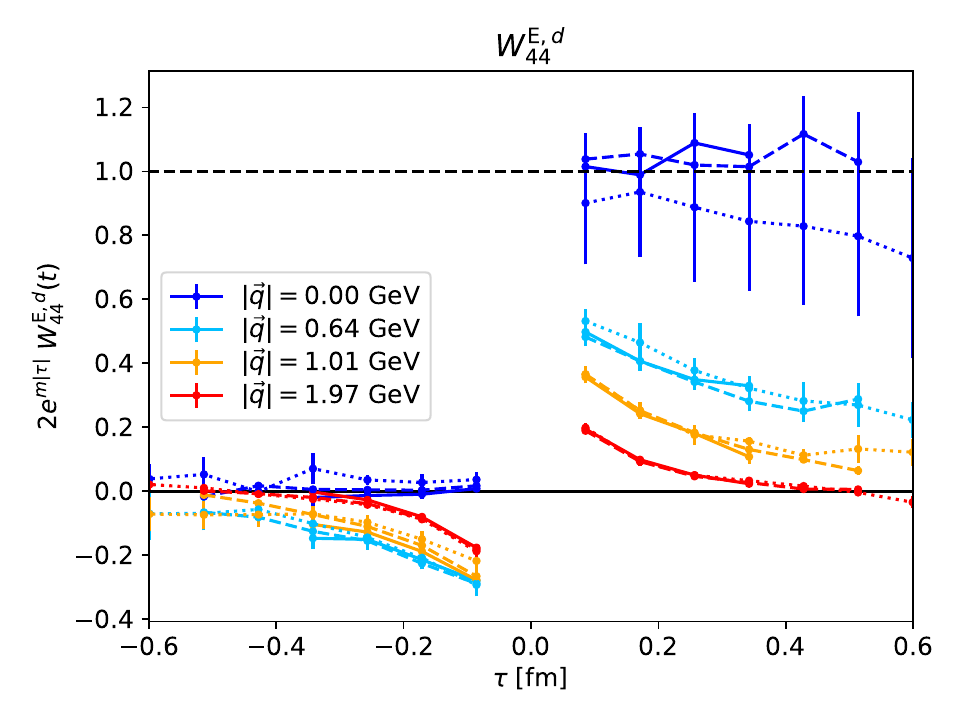}
}
\subfigure[{\parbox[t]{4cm}{$W^{\EE,u}_{44}$, $\tau$-dependence
}\label{fig:W44utau}}]{
\includegraphics[scale=.47, clip, trim=0.4cm 0.5cm 0.4cm 1.15cm]{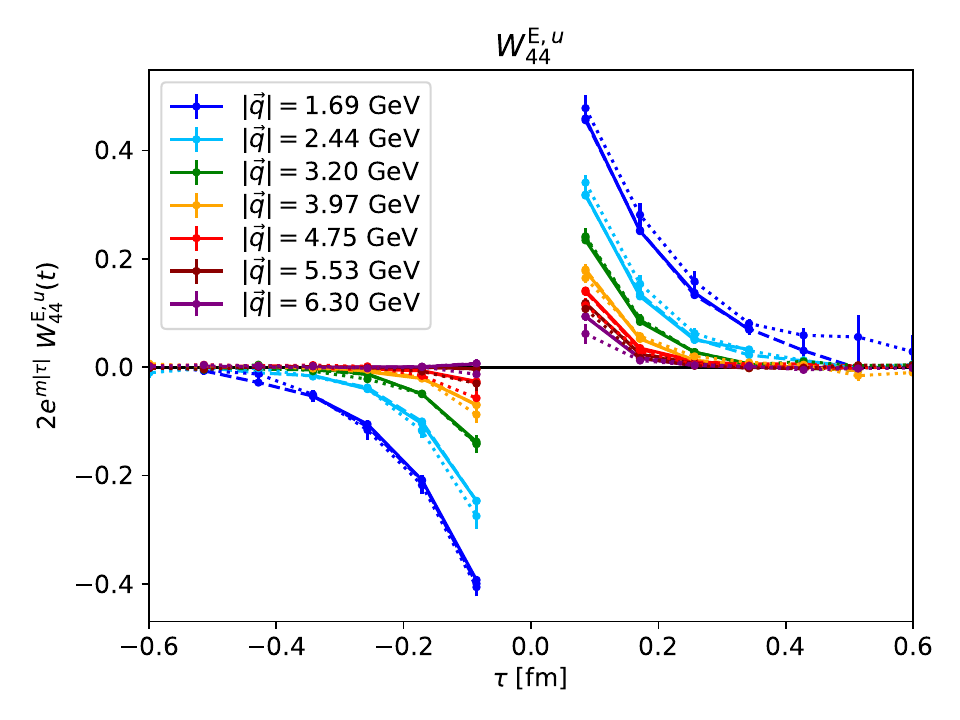}
}
\subfigure[{\parbox[t]{4cm}{$W^{\EE,\mathrm{em}}_{44}$, $\tau$-dependence}\label{fig:W44elmtau}}]{
\includegraphics[scale=.47, clip, trim=0.4cm 0.5cm 0.4cm 1.15cm]{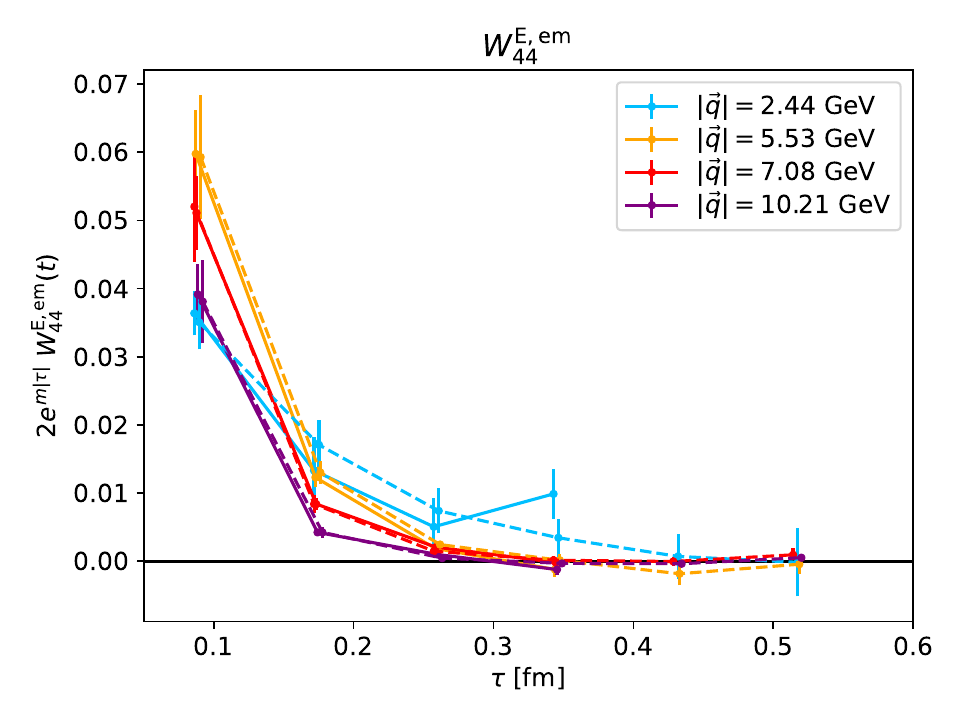}
}
\end{center}
\caption{Panel (a): the dependence of $W^{\EE,u}_{44}$ on $\bar{\tau}$ for $\vec{q}=(2,3,4)2\pi/L$, several source-sink separations $t$, and two different values of $\tau$. Panel (b): the $\tau$-dependence of $W^{\EE,d}_{44}$ for small momentum transfers $\vec{q}$. Panel (c): the $\tau$-dependence of $W^{\EE,d}_{44}$ for larger $\vec{q}$. Panel (d): the $\tau$-dependence of $W^{\EE,\mathrm{em}}_{44}$ for several $\vec{q}$. Panels (b), (c), and (d) show the results for different source-sink separations $t$: solid line: $t=8a$, dashed line: $t=10a$, dotted line: $t=12a$. For better visibility, we applied a small offset w.r.t.\ the horizontal axis for different data sets in panel (d).\label{fig:W44}}
\end{figure}

We start our analysis by looking at the dependence of $W^{\EE}_{44}$ on the average insertion time relative to $t/2$, $\bar{\tau} = (\tau_0+\tau/2)-t/2$, where $\tau_0$ is the timeslice relative to the source of one current, $\tau$ the time distance between the two currents and $t$ the source-sink separation. A potential $\bar{\tau}$-dependence is entirely due to the presence of excited states contaminations. \Fig\ref{fig:W44utaubar} shows the situation for two $u$-quark currents. At the current level of statistics, the data appears to be flat. Moreover it is consistent for all considered source-sink separations. We treat this as an indication that we are only confronted with small contaminations of excited states. 

For further analysis steps, we average the data for $|\bar{\tau}|\le 2a$, while always keeping at least a time distance of $2a$ between each of the two currents and the source or sink, respectively. The resulting values are indicated by the bands in \fig\ref{fig:W44utaubar}. The relevant physics in the context of the hadronic tensor is encoded in the $\tau$-dependence. This is shown in \fig\ref{fig:W44dtau} and \ref{fig:W44utau} for selected momentum transfers $\vec{q}$. Notice that we leave out the data points at $\tau = 0$, which correspond to a contact term. \Fig\ref{fig:W44dtau} includes $\vec{q} = 0$. As a consequence of charge conservation, this is expected to reproduce the number of quarks of the flavor specified by the considered current if $\tau > 0$ (valence quark contribution). Within error bars, this is indeed observed. If $\tau < 0$, we are left with a pure sea-quark contribution, which vanishes at $\vec{q}=\vec{0}$. In particular, the data for $\tau < 0$ corresponds to the \textit{connected} sea \cite{Liu:1993cv}, so that it yields valuable input in the context of studying the Gottfried sum-rule violation \cite{Liu:2020okp}.

Non-zero momentum transfers $\vec{q}\neq\vec{0}$ induce resonances leading to an exponential decay of the the signal, which is clearly visible in our data. Notice that the quality of the signal is significantly enhanced for larger $\vec{q}$ which is a consequence of averaging along different directions of $\vec{q}$ ($W^{\EE}_{44}$ is rotationally invariant in the continuum, we average within the scope of $H_4$-symmetry).

In order to obtain the complete physical contribution to $W^{\EE}_{\mu\nu}$, we have to consider a sum of contractions according to \eqref{eq:proton-elm-wick}. For electromagnetic currents \eqref{eq:Jelm} in a proton, the contribution by the $C_2$-contraction in momentum space reads:

\begin{align}
W^{\EE,\mathrm{em}}_{\mu\nu}(\vec{p},\vec{q},\tau)
=
\frac{4}{9} \left( 
C^{\mu\nu}_{2,u}(\vec{p},\vec{q},\tau) + C^{\nu\mu}_{2,u}(\vec{p},-\vec{q},-\tau) 
\right)
+
\frac{1}{9} \left(
C^{\mu\nu}_{2,d}(\vec{p},\vec{q},\tau) + C^{\nu\mu}_{2,d}(\vec{p},-\vec{q},-\tau) 
\right) \,.
\end{align}
In particular, this requires the sum the contributions of $C_2$ for negative and positive $\tau$.
The corresponding result is plotted in \fig\ref{fig:W44elmtau} for the $44$-component and selected $\vec{q}$. 

\begin{figure}
\begin{center}
\subfigure[{\parbox[t]{4cm}{$A$, $|\vec{q}|$-dependence}\label{fig:Atld}}]{
\includegraphics[scale=.47, clip, trim=0.4cm 0.5cm 0.4cm 1.05cm]{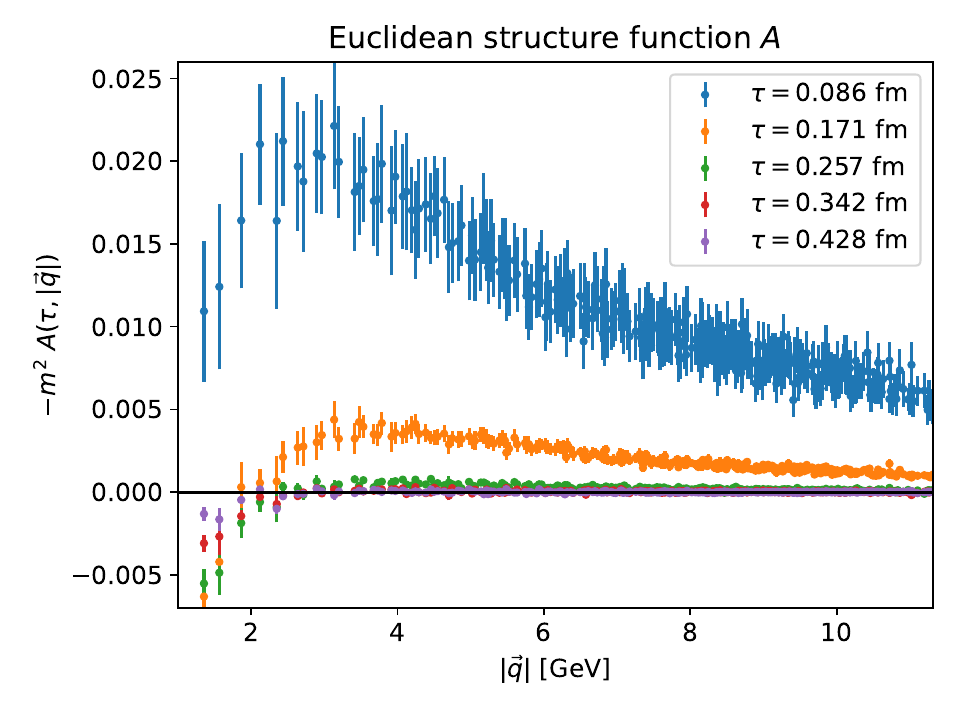}
}
\subfigure[{\parbox[t]{4cm}{$B$, $|\vec{q}|$-dependence}\label{fig:Btld}}]{
\includegraphics[scale=.47, clip, trim=0.4cm 0.5cm 0.4cm 1.05cm]{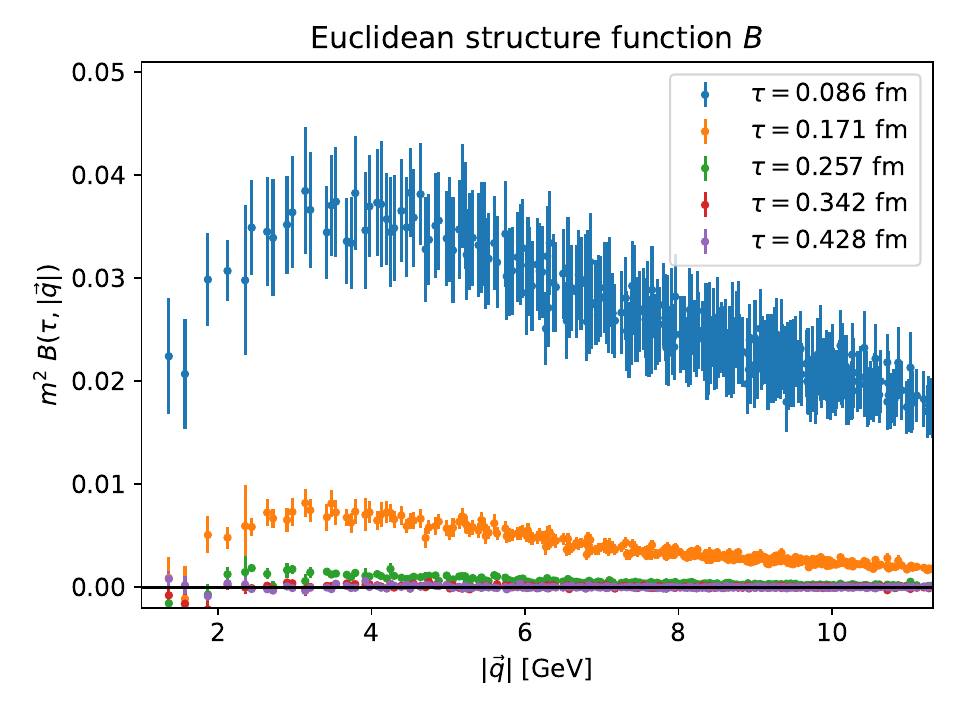}
}
\end{center}
\caption{$|\vec{q}|$-dependence of the Euclidean structure functions $A$ (a) and $B$ (b) for different values of $\tau$ and $t=10a$. The results are shown for the case of electromagnetic currents in an unpolarized proton. \label{fig:ESF}}
\end{figure}
The data for $W^{\EE}_{\mu\nu}$ represents the base for the extraction of the Euclidean structure functions $A$ and $B$ defined in \eqref{eq:WEesys}. These are obtained by solving the over-determined system of equations given by \eqref{eq:WEesys} and \eqref{eq:Ward}. The derivatives $\partial/\partial\tau$ are realized by the symmetric differential quotient w.r.t.\ $\tau$. In order to reduce effects by discretization in the context of derivatives, we interpolate the data of $W^{\EE}_{\mu\nu}(\vec{p},\vec{q},\tau)$ w.r.t.\ $\tau$ using two intermediate steps between each timeslice. Notice that the solutions of $A$ and $B$ for these intermediate timeslices are not considered in the final result. \Fig\ref{fig:ESF} shows the resulting values of $A$ and $B$ as a function of $\vec{q}$ for different values of $\tau$ (different colors) and source-sink separation $t=10a$. One can notice a very fast decay of the signal along $\tau$ so that the signal is consistent with zero for $\tau > 0.3~\mathrm{fm}$ and $|\vec{q}|>3~\mathrm{GeV}$. In order to increase the number of data points along $\tau$, we plan to extend the simulation to smaller lattice spacings.

\section{Conclusion}

We calculated two-current matrix elements on the lattice in order to extract the DIS structure functions using clover fermions. Currently, our analysis includes only one ensemble and $\vec{p}=\vec{0}$. For our purpose, the data has reasonable quality and excited state contaminations appear to be moderate. The current restrictions regarding the nucleon momentum $\vec{p}$ make attempts to deal with the inverse problem for the DIS structure functions unfeasible. Therefore, our next step will be to consider $\vec{p}\neq\vec{0}$. Moreover, we want to extend our analysis to the remaining Wick contractions, in particular $C_1$ (connected), which will be important for anything besides electromagnetic currents in a proton, and $S_2$ (leading disconnected). 
Furthermore, we want to consider axial vectors currents, which play an important role in the context of neutrino-nucleon scattering. We plan to extend our analysis to finer lattices in order to get a better resolution of the $\tau$-dependence of $W^{\EE}_{\mu\nu}$.

\section*{Acknowledgments}

We thankfully acknowledge the CLS collaboration for providing their gauge ensembles.
This work is supported in part by the U.S. Department of Energy, Office of Science, Office of Nuclear Physics, under Grant No.\ DE-SC0013065.
The work of C.Z.\ is supported by the Alexander von Humboldt Foundation.
R.S.S.\ is supported by Laboratory Directed Research and Development (LDRD No. 23-051) of BNL and RIKEN-BNL Research Center.
The authors acknowledge partial support by the U.S. Department of Energy, Office of Science, Office of Nuclear Physics under the umbrella of the Quark-Gluon Tomography (QGT) Topical Collaboration with Award No.\ DE-SC0023646. 
The work was supported in part by the U.S. Department of Energy, Office of Science, Office of Nuclear Physics, under Contract No.\ DE-AC02-05CH11231 that is used to operate Lawrence Berkeley National Laboratory. 
This research used resources of the National Energy Research Scientific Computing Center (NERSC), a U.S. Department of Energy Office of Science User Facility located at Lawrence Berkeley National Laboratory, operated under Contract No.\ DE-AC02-05CH11231. 
We acknowledge the facilities of the USQCD collaboration used for this
research in part, which are funded by the Office of Science of the U.S. Department of Energy.



\end{document}